\documentclass[conference]{IEEEtran}
\IEEEoverridecommandlockouts
\usepackage{cite}
\usepackage{pgfplots}
\usepackage{pgf-pie}
\usepackage{pgfkeys}
\usepackage{amsmath,amssymb,amsfonts}
\usepackage{algorithmic}
\usepackage{graphicx}
\usepackage{textcomp}
\usepackage{xcolor}
\usepackage{url} 
\usepackage{hyperref} 
\def\BibTeX{{\rm B\kern-.05em{\sc i\kern-.025em b}\kern-.08em
    T\kern-.1667em\lower.7ex\hbox{E}\kern-.125emX}}
\usepackage{framed}
\usepackage{lipsum}
\renewenvironment{shaded}{%
\MakeFramed{\advance\hsize-\width \FrameRestore\FrameRestore}}%
{\endMakeFramed}
\definecolor{shadecolor}{gray}{0.75}

\begin{document}
\title{Systems-of-Systems for Environmental Sustainability: A Systematic Mapping Study}

\title{Systems-of-Systems for Environmental Sustainability: A Systematic Mapping Study}

\author{
    \IEEEauthorblockN{Ana Clara Araújo Gomes da Silva}
    \IEEEauthorblockA{\textit{State University of Goiás \& Federal University of Goiás}\\
    Santa Helena de Goiás, Brazil \\
    anaclara.araujo@ueg.br}
    \and
    \IEEEauthorblockN{Gilmar Teixeira Junior}
    \IEEEauthorblockA{\textit{State University of Goiás \& Federal University of Goiás}\\
    Santa Helena de Goiás, Brazil \\
    gilmar.junior@ueg.br}
    \and
    \IEEEauthorblockN{Lívia Mancine C. de Campos}
    \IEEEauthorblockA{\textit \textit{Goiano Federal Institute \& Federal University of Goiás}\\
    Goiânia, Brazil \\
    livia.mancine@ifgoiano.edu.br}
    \and
    \IEEEauthorblockN{Renato F. Bulcão-Neto}
    \IEEEauthorblockA{\textit{Federal University of Goiás}\\
    Goiânia, Brazil \\
    rbulcao@ufg.br}
    \and
    \IEEEauthorblockN{Valdemar Vicente Graciano Neto}
    \IEEEauthorblockA{\textit{Federal University of Goiás}\\
    Goiânia, Brazil \\
    valdemarneto@ufg.br}
}

\maketitle

\begin{abstract}


Environmental sustainability in Systems-of-Systems (SoS) is an emerging field that seeks to integrate technological solutions to promote the efficient management of natural resources. While systematic reviews address sustainability in the context of Smart Cities (a category of SoS), a systematic study synthesizing the existing knowledge on environmental sustainability applied to SoS in general does not exist. Although literature includes other types of sustainability, such as financial and social, this study focuses on environmental sustainability, analyzing how SoS contribute to sustainable practices such as carbon emission reduction, energy efficiency, and biodiversity conservation. We conducted a Systematic Mapping Study to identify the application domains of SoS in sustainability, the challenges faced, and research opportunities. We planned and executed a research protocol including an automated search over four scientific databases. Of 926 studies retrieved, we selected, analyzed, and reported the results of 39 relevant studies. Our findings reveal that most studies focus on Smart Cities and Smart Grids, while applications such as sustainable agriculture and wildfire prevention are less explored. We identified challenges such as system interoperability, scalability, and data governance. Finally, we propose future research directions for SoS and environmental sustainability.

\end{abstract}


\section{Introduction}

Environmental sustainability has been a central theme in science and public policy formulation. Environmental challenges such as climate change, deforestation, and pollution have driven this global agenda, leading to the development of environmental policies, international agreements, and innovative technological solutions \cite{rockstrom2009planetary, steffen2015planetary}. To mitigate these impacts, innovative technological solutions are essential, and Systems-of-Systems (SoS) emerge as a promising approach, enabling the integration of multiple independent systems to achieve common goals. \cite{maier1998architecting} 

Although SoS have been widely studied in domains such as \textit{Smart Cities}, \textit{Smart Grids}, and ecosystem management \cite{jamshidi2017systems}, their relationship with environmental sustainability remains fragmented and underexplored \cite{do2021context}. To bridge this gap, this study conducts a Systematic Mapping Study (SMS) to analyze the role of SoS in natural resource management and the promotion of sustainable practices. The focus is on constituent systems, such as drones and cyber-physical systems, applied to environmental conservation, wildfire prevention, and biodiversity preservation.

The research follows rigorous protocols for selection, extraction, and data synthesis \cite{kitchenham2007guidelines}, allowing for the identification of patterns and trends in the use of SoS for sustainability. The remainder of this study is organized as follows: Section II provides an overview of SoS and Sustainability; Section III details the research method; Section IV discusses the results; and Section V concludes the study, addressing limitations and future research directions.

\section{SoS and Sustainability}


SoS can be classified into different subtypes depending on the level of autonomy and coordination among their constituent systems \cite{maier1998architecting}. The most relevant subtypes include Directed SoS, which has centralized management, Acknowledged SoS, which allows limited autonomy among constituents, and Collaborative SoS, where independent systems interact voluntarily. In environmental applications, these classifications influence how sustainability-oriented SoS are structured and managed.

Emerging technologies such as the Internet of Things (IoT) and Artificial Intelligence (AI) play a central role in enabling these systems \cite{lytras2023digital}, allowing real-time environmental data collection and analysis \cite{lee2015cyber}.  Additionally, they enable early disaster detection, optimization of water and energy resource usage, and support for decision-making in environmental policies \cite{bacco2017environmental}. The literature presents several successful cases of SoS applications for sustainability. One example is the implementation of Smart Grids, which integrate multiple renewable energy sources into intelligent electrical networks, optimizing electricity supply and consumption \cite{fang2012smart}. These networks enable more efficient energy distribution management, reducing waste and contributing to the decarbonization of the electricity sector \cite{bera2025advancing}.



SoS play a key role in biodiversity conservation by integrating multiple independent systems to monitor ecosystems and support conservation strategies. These systems enable large-scale environmental observation, allowing for improved decision-making in habitat preservation and resource management. The use of remote sensing, automated data collection, and real-time analytics enhances biodiversity protection, helping to mitigate threats such as deforestation and habitat degradation.

Studies suggest that the synergy between SoS and sustainability has been widely discussed as a promising approach to addressing complex environmental challenges \cite{adler2024engineering, david2024circular}. However, research indicates that the effective adoption of these systems depends on overcoming technical and institutional barriers, including interoperability, cybersecurity, and data governance \cite{fang2011smart, chabot2013small}. Additionally, the literature highlights that advancing SoS for sustainability requires closer collaboration between governments, industries, and the scientific community, promoting integrated strategies to enable scalable and resilient solutions \cite{hodgson2016best}.

\section{Mapping Method}

This study was conducted under the principles of systematic literature studies, following the guidelines of a Systematic Mapping Study (SMS) as established by Kitchenham and Charters \cite{kitchenham2007guidelines}.




\subsection{Step 1: Planning}

\textit{The study objective was formulated using the GQM (Goal Question Metric) approach proposed by Basili and Weiss \cite{basili_weiss}.} The goal of this study was to \textit{\textbf{analyze} examples of SoS applied to environmental sustainability as reported in the specialized literature, \textbf{with the purpose of} characterizing them \textbf{from the perspective of} researchers \textbf{in the context of} sustainable natural resource management and mitigation of environmental impacts}. This objective led to the formulation of five research questions (RQs) and one research sub-question (RSQ):


\textbf{RQ01: What SoS domain is reported in the study?} \textit{Justification: Specifying the SoS domain addressed, such as Smart Cities, Smart Grids, or Smart Farming, is essential for mapping the study's application areas, facilitating the analysis of similarities and differences between domains and their respective needs.} 

\textbf{RSQ1.1: Within the identified domain, which SoS subtype is best suited for natural resource management?} \textit{Justification: Mapping the specific characteristics that make certain systems more effective in this context facilitates their replication and adaptation in other areas.}



\textbf{RQ02: What are the main challenges faced?}  
\textit{Justification: Analyzing the challenges and opportunities of integrating SoS for sustainability allows us to understand the barriers encountered and the potential benefits of implementing these systems on a large scale, guiding future research and public policies.}  

\textbf{RQ03: What types of constituents are reported?}  
\textit{Justification: Understanding the components that make up the SoS is essential for assessing system complexity, interoperability, and the specific roles of each element in achieving the overall system goals.}  

\textbf{RQ04: What are the main functionalities of the reported constituent systems?}  
\textit{Justification: Mapping how systems collaborate to achieve the overall objectives of the SoS also allows for assessing how these functionalities can be optimized or replicated in other contexts.}

\textbf{RQ05: What are the missions and emerging behaviors?}
\textit{Justification: Understanding how systems interact and what unexpected or innovative outcomes may arise from these interactions is crucial. This provides insights into the adaptability and overall impact of the SoS.}

\subsection{Step 2: Mapping Protocol}
The SMS followed a rigorous protocol composed of the following steps: (i) definition of the search string, (ii) selection of databases, (iii) application of inclusion (CI) and exclusion (CE) criteria, (iv) selection process, and (v) data extraction and analysis. For conducting the study, we used the web tool Parsif.al\footnote{\url{https://parsif.al/}}.

\subsubsection{Search String}
The search strategy was defined based on previous studies \cite{webster2002analyzing}, ensuring comprehensive coverage of relevant studies. The adopted search string was structured to capture publications that relate SoS to environmental sustainability. The search string used was:

\textit{``System of Systems" OR  ``Systems of Systems" OR  ``System-of-system" OR  ``systems-of-systems" OR  ``Systems of Autonomous Systems" OR  ``CyberPhysical Systems-of-Systems" OR   ``Systems of Information Systems") AND (Sustainab* OR  Green).} 

\subsubsection{Consulted Databases}

The main scientific databases were consulted, ensuring reproducibility and broad literature coverage: IEEE Xplore, Scopus, Web of Science, and ACM, which are among the key sources for conducting systematic studies \cite{guessi2015systematic}.

\subsubsection{Inclusion and Exclusion Criteria}

The studies were selected based on the following criteria:

\textbf{Inclusion Criterion:} \textbf{CI1} The primary study explicitly addresses \textbf{SoS and sustainability}.

\textbf{Exclusion Criteria:} \textbf{CE1} The study \textbf{is not a primary study} (e.g., literature reviews or conceptual studies without practical application), \textbf{CE2} The study \textbf{is not peer-reviewed} (gray literature), \textbf{CE3} The study \textbf{is not in English}, \textbf{CE4} The study \textbf{does not address SoS}, \textbf{CE5} The study \textbf{does not address sustainability}, \textbf{CE6} The study \textbf{is a duplicate or an earlier version of an already updated and included study}.

\subsubsection{Selection Process}

The selection process was conducted using the Parsif.al tool. In this process, two researchers independently reviewed the studies and decided whether to include or exclude the article. In cases of uncertainty, the article was annotated with the word “doubt.” In such cases, a third researcher with more experience conducted an additional review to determine whether to include or exclude the article. Whenever disagreements arose, the researchers met to discuss and reach a consensus. To ensure the replicability and transparency of the results, details of the process are available on \textit{Zenodo} \footnote{Available at: \url{https://doi.org/10.5281/zenodo.14807834}}. The steps followed were:

\begin{itemize}
    \item[] \textbf{Search Execution}: 
    The search string was applied to titles, abstracts, and keywords in the bibliographic databases and search engines selected in January 2025. As a result, 923 studies were identified, distributed as follows: 48 in ACM, 211 in IEEE, 510 in Scopus, and 157 in Web of Science.

    \item[] \textbf{Duplicate Removal}:
    A total of 117 duplicate studies were eliminated, leaving 810 studies for analysis.

    \item[] \textbf{Filtering Phases}: 
    The study selection was performed in three stages: \textbf{(i)} Analysis of study titles and abstracts. According to the exclusion criteria, 304 studies were excluded, leaving 506 studies. \textbf{(ii)} Analysis of the introduction and conclusion of the remaining studies, leading to the exclusion of additional studies. \textbf{(iii)} Full-text reading, which resulted in the selection of \underline{\textbf{39 studies}} to form the final mapping database.

\end{itemize}

\subsubsection{Data Extraction and Analysis}

The data from the \textbf{\underline{studies}} were systematically extracted, collecting demographic and specific details from the studies. The fields included in the extraction form (EF) are listed in Table~\ref{tab:extracao_dados}.

\begin{table}[h]
\centering
\footnotesize
\caption{Fields of the Data Extraction Form}
\label{tab:extracao_dados}
\begin{tabular}{|c|p{6cm}|} 
\hline
\textbf{Code} & \multicolumn{1}{c|}{\textbf{Field Description}} \\
\hline
EF1 - EF5 & Basic study metadata (title, publication year, authors, publication venue, and DOI/URL). \\
\hline
EF6 & Classification of the study type. \\
\hline
EF7 & Type of sustainability addressed in the study. \\
\hline
EF8 & SoS domains reported in the study. \\
\hline
EF9 & SoS subtype applied to natural resource management. \\
\hline
EF10 & Main challenges faced. \\
\hline
EF11 & Types of constituents reported in the study. \\
\hline
EF12 & Main functionalities of the constituent systems. \\
\hline
EF13 & Missions and emerging behaviors. \\
\hline
\end{tabular}
\end{table}

The EF fields were aligned with the research questions (RQs and SRQs) to ensure consistency and relevance. Table \ref{tabela1} illustrates the mapping of research questions to extraction form fields.

\begin{table}[h!]
\centering
\caption{Relationship Between Research Questions and EF Questions}
\begin{tabular}{|c|c|}
\hline
\textbf{Research Question} & \textbf{EF Question} \\ \hline
\textbf{RQ01} & EF8 \\ \hline
\textbf{SRQ1.1} & EF9 \\ \hline
\textbf{RQ02} & EF10 \\ \hline
\textbf{RQ03} & EF11 \\ \hline
\textbf{RQ04} & EF12 \\ \hline
\textbf{RQ05} & EF13 \\ \hline
\end{tabular}
\label{tabela1}
\end{table}

Data extraction was conducted considering methodological aspects and results presented in the studies. The analysis followed a quantitative approach \cite{petersen2015guidelines}, categorizing the contributions of the studies to sustainability using SoS.

\section{Results Analysis}

This section presents the results of the analysis of the selected studies. Due to space constraints, it was not possible to discuss each study in detail. However, since all 39 included studies have been cited throughout the text, we provide a summarized reference table (\autoref{tab:articles}) to explicitly list them. The complete dataset, including additional metadata, is available on \textit{Zenodo} \footnote{Available at: \url{https://doi.org/10.5281/zenodo.14807834}}.

\begin{table}[h!]
\centering
\caption{List of Included Studies}
\begin{tabular}{|p{8cm}|}
\hline
\multicolumn{1}{|c|}{\textbf{Included Studies}} \\ \hline
\cite{hipel2013tackling}, \cite{hall2014water}, \cite{liu2016reengineering}, \cite{fritz2008conceptual}, \cite{geldenhuys2018literature}, \cite{teymourifar2023using}, \cite{sebestyen2021applicability}, \cite{sheridan2024contextualising}, \cite{pastor2024r}, \cite{boldrini2022multi}, \cite{thatcher2024applying}, \cite{yap2024application}, \cite{heitmann2019requirements}, \cite{guan2023scalable}, \cite{arnautovic2012value}, \cite{leach2019liveable}, \cite{lai2022urban}, \cite{vercruysse2019interoperability}, \cite{delgado2019big}, \cite{elnashai2021vision}, \cite{howell2017towards}, \cite{rogers2023delivering}, \cite{chang2024climate}, \cite{beckett2023pathway}, \cite{albaaji2023artificial}, \cite{prakashacolossus}, \cite{markey2023managing}, \cite{nozarian2024system}, \cite{abdoli2021analyzing}, \cite{watson2020systems}, \cite{little2019tiered}, \cite{tariq2013determinants}, \cite{hoffmann2014improved}, \cite{hernandez2014multi}, \cite{phillis2012system2}, \cite{sandor2008transitioning}, \cite{scholes2012building}, \cite{phillis2012system}, \cite{pruyt2007transition} \\ 
\hline
\end{tabular}
\label{tab:articles}
\end{table}



\subsection{Which SoS domain is reported in the study? (RQ01)}

\begin{figure}[h]
    \centering
    \includegraphics[width=\columnwidth]{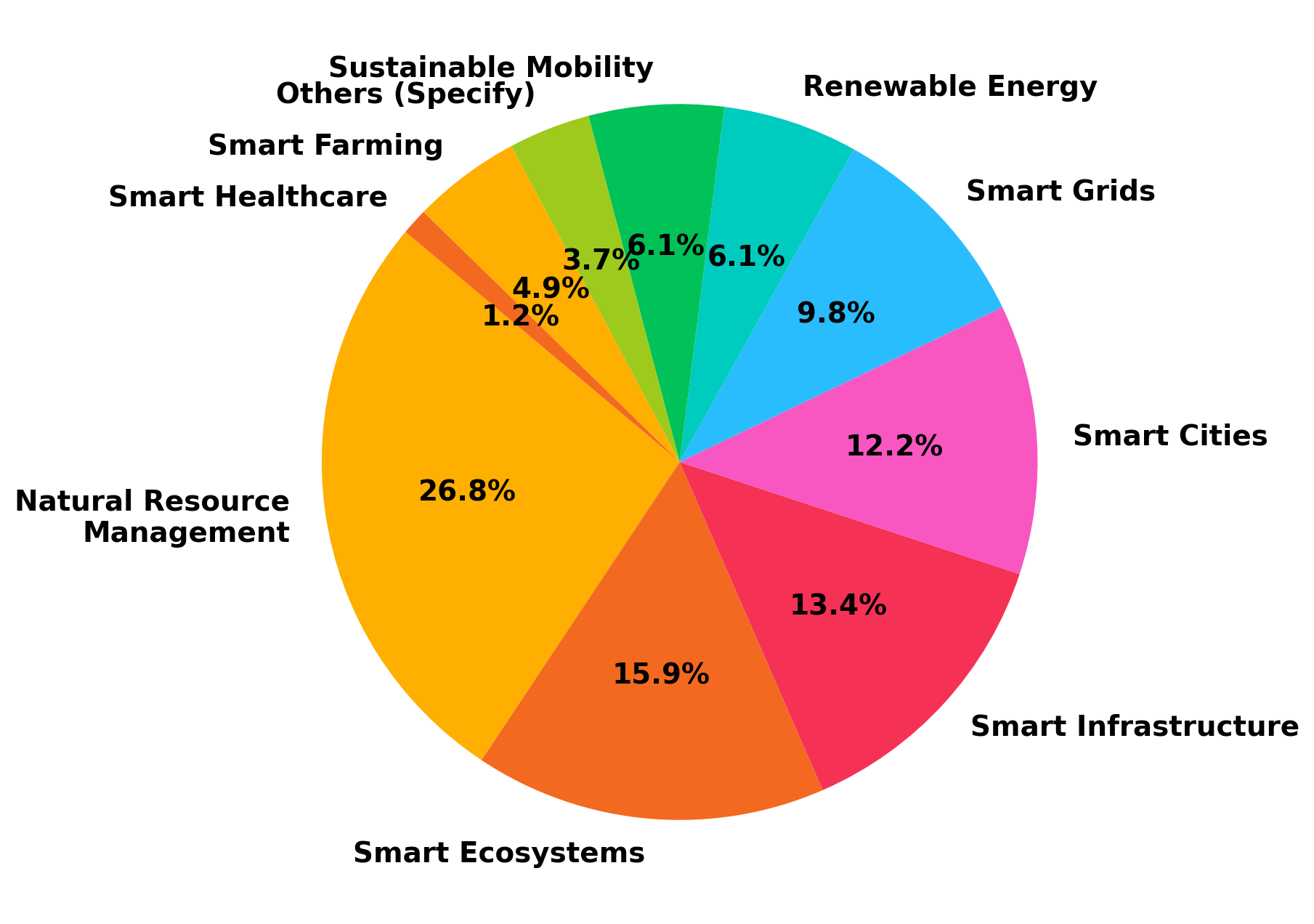}
    \caption{SoS domains reported in the studies}
    \label{fig1}
\end{figure}

The reviewed studies highlight \textbf{Natural Resource Management} (26.8\%) as the most explored domain, followed by \textit{Smart Ecosystems} (15.9\%) and \textit{Smart Infrastructure} (13.4\%). Additionally, \textbf{Smart Cities} and \textbf{Smart Grids} also stand out, as illustrated in Figure \ref{fig1}.

In the context of \textbf{Smart Cities}, studies such as \cite{geldenhuys2018literature}, \cite{lai2022urban}, and \cite{nozarian2024system} discuss the transition toward smarter and more sustainable cities, emphasizing the role of urban infrastructure integration and data-driven approaches to support carbon neutrality. Additionally, research like \cite{elnashai2021vision} and \cite{rogers2023delivering} highlights innovative strategies to make cities more resilient and livable, promoting transformative governance and data-driven urban planning. 

In the context of \textbf{Smart Grids}, studies such as \cite{howell2017towards} and \cite{tariq2013determinants} explore the application of multi-agent models and intelligent demand-side management to optimize energy consumption and distribution. Furthermore, \cite{little2019tiered} presents a hierarchical decision-making model to address energy and socio-environmental challenges in an integrated manner.


Sustainable agriculture and wildfire prevention studies were also identified, though in smaller numbers. In \textbf{sustainable agriculture}, studies like \cite{delgado2019big} and \cite{albaaji2023artificial} highlight the use of IoT sensors, machine learning, and drones to optimize natural resource management, enabling automated soil monitoring, irrigation control, and crop prediction based on climate data. However, the characterization of sensors as constituents in these systems can be questioned. Additionally, \cite{guan2023scalable} proposes a model for quantifying agricultural carbon emissions, using large-scale data analysis to map environmental impacts and suggest more sustainable practices, contributing to climate change mitigation in the sector.

Despite the identified applications of SoS in wildfire prevention, there are still open challenges and unexplored opportunities in the field of environmental sustainability. For instance, no studies were found that specifically address the use of SoS for riparian forest restoration, ensuring water conservation in river ecosystems. Similarly, while some studies discuss pollution monitoring, the implementation of SoS for proactive pollution prevention remains underexplored. Other potential areas for development include SoS-based solutions for waste management, air humidity regulation, and large-scale environmental remediation. These gaps highlight the need for further research to expand the role of SoS in addressing key environmental challenges beyond those currently explored in the literature.

\begin{shaded}
\textbf{Finding 1:} Most studies focus on \textbf{Natural Resource Management} (26.8\%), \textbf{Smart Ecosystems} (15.9\%), and \textbf{Smart Infrastructure} (13.4\%), with additional emphasis on \textbf{Smart Cities} and \textbf{Smart Grids}. However, key environmental applications, such as \textbf{sustainable agriculture}, \textbf{wildfire prevention}, and \textbf{pollution prevention}, remain underexplored, highlighting the need to broaden SoS research in sustainability.
\end{shaded}

\subsection{Within the identified domain, which SoS subtype best applies to natural resource management? Which SoS domain is reported in the study? (SRQ01)}

The analysis revealed that most studies (22 out of 39, 57.5\%) adopt an \textit{Acknowledged} SoS approach, where independent systems operate with common objectives, such as optimizing energy management and environmental monitoring. Additionally, 12 studies (30\%) explore \textit{Collaborative} SoS, mainly in research focused on climate data integration and environmental sensor networks. Finally, 5 studies (12.5\%) address \textit{Directed} SoS, in which centralized coordination manages the interaction between constituent systems.

By cross-referencing this information with the domains identified in SRQ01, it is observed that Acknowledged SoS predominates in Smart Grids and Smart Cities, where independent systems cooperate for urban infrastructure management and energy efficiency. Collaborative SoS is more common in studies on environmental monitoring and climate management, as it enables decentralized data sharing among different autonomous systems. Meanwhile, Directed SoS tends to be used in emergency response and disaster management, such as wildfire detection and control, where centralized control is essential for rapid decision-making.

\subsection{What are the main challenges faced? (RSQ02)}

The most frequently cited challenges in the studies include: \textbf{i) Interoperability between different systems}: difficulty in integrating heterogeneous platforms and sensors, \textbf{ii) Governance and regulation}: lack of standardization in environmental information exchange, \textbf{iii) Solution scalability}: adaptation of technologies to different regions and climate contexts, and \textbf{iv) Environmental data management}: the need for robust infrastructure for large-scale data collection and analysis.

Interoperability between independent systems was identified as a critical challenge, as discussed in \cite{vercruysse2019interoperability}, which addresses the need for data integration for urban risk management. Additionally, \cite{heitmann2019requirements} emphasizes regulation for the secure sharing of environmental data among multiple \textit{stakeholders}, a problem also raised by \cite{hoffmann2014improved}, which highlights interoperability and data-sharing difficulties for large-scale biodiversity monitoring.

\begin{shaded}
\textit{\textbf{Finding 2:}} \textit{Key challenges in applying SoS to sustainability include (i) interoperability between heterogeneous systems, (ii) governance of environmental data, (iii) scalability of solutions, and (iv) regulatory barriers. These obstacles hinder large-scale implementation and require multidisciplinary collaboration for effective solutions.}
\end{shaded}

\subsection{What are the types of constituents? (RQ03)}

The constituents of environmental SoS vary according to the application domain. The most recurrent ones include: distributed sensors for climate and energy monitoring \cite{delgado2019big} \cite{boldrini2022multi} \cite{nozarian2024system} \cite{teymourifar2023using} intelligent urban infrastructures such as transportation networks and electrical systems \cite{lai2022urban} \cite{howell2017towards} \cite{liu2016reengineering} \cite{rogers2023delivering}. The use of UAVs (drones) for remote environmental monitoring and emergency response \cite{albaaji2023artificial} \cite{little2019tiered} \cite{sebestyen2021applicability}, including wildfire prevention and mitigation, is discussed in \cite{pastor2024r} \cite{albaaji2023artificial}.

\begin{shaded}
\textit{\textbf{Finding 3:}} \textit{The primary constituents of environmental SoS include (i) distributed sensors for environmental monitoring, (ii) drones for remote sensing, intelligent urban infrastructure, and multi-agent systems for decision-making. These elements collectively enable real-time analysis and adaptive responses.}
\end{shaded}

\subsection{What are the main functionalities of the reported constituent systems? (RQ04)}

The most frequently mentioned functionalities in the analyzed studies include: Environmental monitoring (climate sensors and observation networks); Prediction and modeling (simulation of future scenarios); Resource management (optimization of water and energy use); Platform integration (interconnectivity between different systems); Automation and control (automatic responses to environmental changes). The Table \ref{tab:funcionalidades_estudos} presents the functionalities and the studies that address them

\begin{table}[h]
    \centering
    \caption{Functionalities and Related Studies}
    \label{tab:funcionalidades_estudos}
    \renewcommand{\arraystretch}{1.2}
    \resizebox{\columnwidth}{!}{ 
    \begin{tabular}{|p{3.5cm}|p{5.5cm}|c|}
        \hline
        \textbf{Functionality} & \textbf{Studies} & \textbf{Frequency} \\
        \hline
        Monitoring (environmental data capture, surveillance of specific areas, heat source detection) & \cite{hipel2013tackling}, \cite{hall2014water}, \cite{liu2016reengineering}, \cite{fritz2008conceptual}, \cite{geldenhuys2018literature}, \cite{teymourifar2023using}, \cite{sebestyen2021applicability}, \cite{sheridan2024contextualising}, \cite{boldrini2022multi}, \cite{thatcher2024applying}, \cite{heitmann2019requirements}, \cite{guan2023scalable}, \cite{arnautovic2012value}, \cite{pastor2024r}, \cite{vercruysse2019interoperability}, \cite{delgado2019big}, \cite{elnashai2021vision}, \cite{howell2017towards}, \cite{lai2022urban}, \cite{chang2024climate}, \cite{albaaji2023artificial}, \cite{prakashacolossus}, \cite{markey2023managing}, \cite{nozarian2024system}, \cite{abdoli2021analyzing}, \cite{little2019tiered}, \cite{hernandez2014multi}, \cite{tariq2013determinants},  \cite{scholes2012building}, \cite{phillis2012system}, \cite{hoffmann2014improved}, \cite{phillis2012system2} & 32 \\
        \hline
       Data Analysis (mapping of vulnerable areas, prediction of critical conditions, pattern identification) & \cite{hipel2013tackling}, \cite{hall2014water}, \cite{liu2016reengineering}, \cite{teymourifar2023using}, \cite{sebestyen2021applicability}, \cite{heitmann2019requirements}, \cite{chang2024climate}, \cite{lai2022urban}, \cite{delgado2019big}, \cite{pastor2024r}, \cite{prakashacolossus}, \cite{markey2023managing}, \cite{nozarian2024system}, \cite{abdoli2021analyzing}, \cite{watson2020systems}, \cite{little2019tiered}, \cite{hoffmann2014improved}, \cite{hernandez2014multi}, \cite{phillis2012system}, \cite{riley2008transitioning}, \cite{scholes2012building}, \cite{phillis2012system2}, \cite{yap2024application}, \cite{tariq2013determinants} & 24 \\
        \hline
        Forecasting and Modeling (scenario simulation, such as fire spread or future resource demands) & \cite{hall2014water}, \cite{liu2016reengineering}, \cite{teymourifar2023using}, \cite{sebestyen2021applicability}, \cite{thatcher2024applying}, \cite{arnautovic2012value}, \cite{boldrini2022multi}, \cite{guan2023scalable}, \cite{lai2022urban}, \cite{delgado2019big}, \cite{prakashacolossus}, \cite{markey2023managing}, \cite{nozarian2024system}, \cite{abdoli2021analyzing}, \cite{watson2020systems}, \cite{little2019tiered}, \cite{chang2024climate}, \cite{hernandez2014multi}, \cite{phillis2012system}, \cite{scholes2012building}, \cite{phillis2012system2}, \cite{albaaji2023artificial}, \cite{tariq2013determinants}, \cite{sandor2008transitioning}  & 24 \\
        \hline
        Resource Management (optimization of water, energy, and other resource usage) & \cite{hipel2013tackling}, \cite{hall2014water}, \cite{liu2016reengineering}, \cite{fritz2008conceptual}, \cite{geldenhuys2018literature}, \cite{pastor2024r}, \cite{abdoli2021analyzing}, \cite{tariq2013determinants}, \cite{phillis2012system2}, \cite{albaaji2023artificial}, \cite{pruyt2007transition}, \cite{markey2023managing}, \cite{nozarian2024system}, \cite{abdoli2021analyzing}, \cite{watson2020systems}, \cite{little2019tiered}, \cite{hoffmann2014improved}, \cite{hernandez2014multi}, \cite{phillis2012system}, \cite{riley2008transitioning}, \cite{scholes2012building} & 21 \\
        \hline
       Environmental Impact Mitigation (restoration of affected areas, emission control, reforestation) & \cite{hipel2013tackling}, \cite{hall2014water}, \cite{liu2016reengineering}, \cite{heitmann2019requirements}, \cite{vercruysse2019interoperability}, \cite{pastor2024r}, \cite{albaaji2023artificial}, \cite{prakashacolossus}, \cite{markey2023managing}, \cite{nozarian2024system}, \cite{abdoli2021analyzing}, \cite{beckett2023pathway}, \cite{little2019tiered}, \cite{yap2024application}, \cite{phillis2012system}, \cite{phillis2012system2}, \cite{scholes2012building} & 17 \\
        \hline
       Automation (execution of automated actions, such as irrigation or fire suppression) & \cite{hipel2013tackling}, \cite{fritz2008conceptual}, \cite{delgado2019big}, \cite{watson2020systems}, \cite{pastor2024r}, \cite{albaaji2023artificial}, \cite{prakashacolossus}, \cite{markey2023managing}, \cite{little2019tiered}, \cite{riley2008transitioning},  \cite{pruyt2011energy} & 11 \\
        \hline
        Platform Integration & 
        \cite{liu2016reengineering}, \cite{arnautovic2012value}, \cite{elnashai2021vision}, \cite{rogers2023delivering}, \cite{pastor2024r}, \cite{beckett2023pathway}, \cite{markey2023managing}, \cite{prakashacolossus}, \cite{little2019tiered}, \cite{hoffmann2014improved}, \cite{scholes2012building} & 11 \\
        \hline
        Communication (data transmission between constituents, real-time alert delivery) & \cite{liu2016reengineering}, \cite{hall2014water}, \cite{geldenhuys2018literature}, \cite{vercruysse2019interoperability}, \cite{fritz2008conceptual}, \cite{elnashai2021vision}, \cite{prakashacolossus}, \cite{hoffmann2014improved},  & 8 \\
        \hline
        Others (specify) &
        \cite{sheridan2024contextualising}, \cite{boldrini2022multi}, \cite{thatcher2024applying}, \cite{leach2019liveable}, \cite{howell2017towards}, \cite{markey2023managing} & 6 \\
        \hline
        Autonomous Collaboration (coordination between constituents, such as drones and sensors, for joint efforts) &
        \cite{sebestyen2021applicability}, \cite{heitmann2019requirements}, \cite{little2019tiered}, \cite{phillis2012system} & 4 \\
        \hline
        Quality Control and Maintenance (system diagnostics and preventive maintenance) &
        \cite{geldenhuys2018literature}, \cite{heitmann2019requirements} & 2 \\
        \hline
        Critical Infrastructure Management & \cite{hoffmann2014improved}  & 1 \\
        \hline
        \end{tabular}
    }
\end{table}

\begin{shaded}
\textit{\textbf{Finding 4:}} \textit{The most common functionalities identified in SoS for sustainability include (i) environmental monitoring, (ii) predictive modeling, (iii) resource management, (iv) platform integration, and (v) automated control. These capabilities enhance system efficiency and contribute to sustainable natural resource use.}
\end{shaded}


\subsection{What are the missions and emerging behaviors? (RQ05)}

The missions of the analyzed SoS include: (i) Carbon emission reduction and mitigation of environmental impacts; (ii) Promotion of urban and environmental resilience against climate change; (iii) Optimization of sustainable infrastructure management, especially in smart cities.

The most observed emerging behaviors were: (i) Collaboration between independent emergent systems for environmental data sharing; (ii) Dynamic adaptation to extreme climate events; (iii) Use of artificial intelligence to optimize system efficiency.

The concept of missions and emergent behavior in environmental SoS was explored in \cite{pruyt2007transition}, which highlights the growing integration of renewable sources to create decentralized electrical grids. Additionally, \cite{markey2023managing} shows how the use of \textit{Digital Twins} can optimize the management and automation of sustainable infrastructures.

\begin{shaded}
\textit{\textbf{Finding 5:}} \textit{The missions of environmental SoS primarily focus on carbon emission reduction, climate change resilience, and smart infrastructure optimization. Emerging behaviors include increased collaboration among independent systems, dynamic adaptation to environmental conditions, and AI-driven decision-making.}
\end{shaded}


\section{Discussions and Implications}

The results of this systematic mapping reinforce the growing importance of SoS for environmental sustainability, highlighting different approaches and challenges faced by researchers. The predominance of studies focused on \textit{Smart Cities} and \textit{Smart Grids} suggests that these domains are considered priorities for the integration of sustainable solutions, as evidenced by studies such as \cite{lai2022urban} and \cite{howell2017towards}. Each identified domain presents distinct patterns in the composition of its constituents and the predominant type of SoS. For example, Smart Cities use IoT sensors, communication networks, and Big Data platforms, typically organized as an Acknowledged SoS, where there is hierarchical coordination but with partial autonomy of the systems. Smart Grids, on the other hand, operate similarly but focus on energy optimization, using smart meters and multi-agent systems for load balancing.

However, the lower number of studies on sustainable agriculture and ecosystem preservation suggests a research gap to be explored. While the study by \cite{hoffmann2014improved} demonstrates advances in data collection and analysis for biodiversity, few works delve into the use of SoS for climate change mitigation in the agricultural sector. The findings of this study provide direct insights for researchers and professionals interested in implementing SoS for sustainability. For instance, to develop an SoS for water management in Goiás, water quality sensors, IoT networks, and predictive modeling are essential components, organized in a Collaborative SoS. In the Cerrado bioma (a type of Brazilian savanna), where climate variability is a major challenge, solutions based on drones for vegetation analysis, soil moisture sensors, and AI for crop prediction stand out, reinforcing the role of Collaborative SoS in optimizing natural resources.

Another relevant point is the diversity of definitions and applications of the sustainability concept. While most studies address sustainability from an environmental perspective, some research considers broader approaches, including economic and social aspects, as discussed in \cite{rogers2023delivering} and \cite{beckett2023pathway}.

Modeling and simulation approaches emerge as essential tools for capturing the complexity of SoS applied to environmental sustainability. As highlighted by Chang and Hossain \cite{chang2024climate}, computational modeling techniques are fundamental for predicting climate risks and improving the resilience of urban infrastructures, assisting in the formulation of sustainable policies. Beyond the differences in constituents, each domain adopts a specific type of SoS according to its needs. In wildfire prevention, for example, thermal sensors, satellites, and UAVs are widely used, organized into a Directed SoS, in which a central control system manages the emergency response. In sustainable agricultural projects, climate sensors and machine learning platforms operate in a decentralized manner, forming a Collaborative SoS, where autonomous systems share information to optimize the use of natural resources.

Complementarily, Thatcher, Robinson, and Brown \cite{thatcher2024applying} demonstrate how simulation contributes to optimizing water and urban waste management, promoting greater efficiency in resource allocation. Additionally, Gomes Júnior et al. \cite{gomes2022flood} emphasize that digital transformation is driving the evolution of SoS, particularly in domains such as mobility and energy, requiring dynamic modeling approaches to capture the evolving nature of these systems. In this context, Graciano Neto et al. \cite{manzano2020dynamic} have explored innovative approaches to modeling and simulation of SoS in complex scenarios, highlighting the importance of flexible and scalable architectures for assessing the sustainability of intelligent infrastructures. These studies reinforce the need to improve simulation techniques to evaluate future scenarios and support strategic decisions for sustainability.

\subsection{Research Agenda}

Based on the findings of this systematic mapping, we identified several gaps and research opportunities that can guide future investigations on SoS and environmental sustainability. Below, we present a research agenda structured into three main axes:

\begin{itemize}
    \item \textbf{Expanding the Application Domains of SoS for Sustainability} \\
    The predominance of studies focused on Smart Cities and Smart Grids indicates a need to expand the application of SoS to other domains. Sectors such as sustainable agriculture, ecosystem management, and natural disaster prevention remain underexplored. Additionally, domains like water management, reforestation of riparian zones, air humidity control, pollution, and waste disposal were not found in the included studies. Future research can investigate how SoS modeling and simulation can contribute to environmental monitoring, climate impact mitigation, and efficient use of natural resources across different sectors.
   
    \item \textbf{Development of Architectures and Models for SoS Applications in Sustainability} \\
    The interoperability between constituent systems, environmental data governance, and the resilience of sustainable infrastructures are recurring challenges in the literature. New architectural models can be explored to improve communication between autonomous systems, ensure information security, and optimize real-time decision-making. Additionally, approaches based on digital twins and artificial intelligence can strengthen the predictive and adaptive capacity of SoS in environmental contexts.

    \item \textbf{Guidelines for Evaluating and Validating Sustainability Provided by SoS} \\
    The lack of standardized metrics to assess the environmental impact of SoS represents a gap in the literature. Future research may focus on developing specific indicators to measure the effectiveness of SoS in promoting environmental sustainability, including metrics for emission reduction, energy efficiency, and biodiversity preservation. Moreover, experimental studies and real-world applications can contribute to validating the benefits and limitations of SoS in sustainable scenarios.
\end{itemize}

\subsection{Threats to Validity}

Despite the methodological rigor adopted in this systematic mapping, some threats to validity must be considered. Regarding \textbf{internal validity}, although inclusion and exclusion criteria were clearly defined, the possibility of bias in the selection of studies cannot be ruled out. Potentially relevant studies may not have been included due to limitations in the keywords used in the search or the unavailability of the full text. Additionally, while the data extraction process was conducted systematically, there may be subjectivity in the categorization of studies, especially in classifying types of sustainability and the challenges faced by SoS. 

Concerning \textbf{external validity}, the findings of this mapping reflect trends observed in the analyzed literature but may not be generalizable to all SoS and sustainability contexts. Future studies with broader scopes or empirical approaches may complement this analysis. Although this study focused on environmental sustainability, other dimensions (economic and social) are also relevant and may have been underestimated. Future research could explore the integration of these dimensions within SoS.

Finally, regarding \textbf{construct validity}, the literature on SoS and sustainability is constantly evolving. Recent studies may introduce new concepts and approaches that have not yet been covered in this mapping. It is recommended that this study be periodically replicated to track advancements in the field. Another important point is that, although the databases used are widely recognized, some relevant studies may be found in other sources, such as gray literature and emerging publications that were not considered.

Given these threats, we sought to mitigate their impact through rigorous selection and data extraction criteria, as well as a structured analysis of the reviewed studies. However, we acknowledge that future research may complement this investigation by validating or expanding the reported findings.

\section{Final Remarks}

This SMS highlighted the relevance of \textbf{SoS} for environmental sustainability, emphasizing their application domains, challenges, and opportunities. The analyzed studies are primarily focused on \textit{Smart Cities} and \textit{Smart Grids}, while other applications, such as sustainable agriculture and environmental conservation, are less explored despite their potential.

The identified challenges include the need for greater interoperability between systems, data governance, and model standardization, as discussed in \cite{phillis2012system} and \cite{sandor2008transitioning}. Additionally, the diversity of conceptual approaches to sustainability reinforces the need for a clearer alignment between future research.

Based on these analyses, we recommend that future research broaden the scope of SoS applications for sustainable practices, addressing underexplored sectors such as \textbf{agriculture, ecosystem management, and public policies for climate change mitigation}. Furthermore, the integration of \textbf{emerging technologies}, such as \textbf{Artificial Intelligence and Blockchain}, could enhance the efficiency and reliability of SoS in promoting environmental sustainability.

These findings reinforce the importance of continued research on the role of SoS in sustainability, encouraging further investigations into \textbf{integration strategies, governance, and the environmental impact of such systems across different contexts}.

\section*{Acknowledgements}

Valdemar V. Graciano Neto thanks the Goiás Research Support Agency (FAPEG) for partially covering the expenses to develop and present this work under public call No. 11/2024, grant number 2024/10267001457.

\bibliographystyle{IEEEtran}
\bibliography{IEEEabrv,IEEEexample}

\begin{thebibliography}{10}
\providecommand{\url}[1]{#1}
\csname url@samestyle\endcsname
\providecommand{\newblock}{\relax}
\providecommand{\bibinfo}[2]{#2}
\providecommand{\BIBentrySTDinterwordspacing}{\spaceskip=0pt\relax}
\providecommand{\BIBentryALTinterwordstretchfactor}{4}
\providecommand{\BIBentryALTinterwordspacing}{\spaceskip=\fontdimen2\font plus
\BIBentryALTinterwordstretchfactor\fontdimen3\font minus \fontdimen4\font\relax}
\providecommand{\BIBforeignlanguage}[2]{{%
\expandafter\ifx\csname l@#1\endcsname\relax
\typeout{** WARNING: IEEEtran.bst: No hyphenation pattern has been}%
\typeout{** loaded for the language `#1'. Using the pattern for}%
\typeout{** the default language instead.}%
\else
\language=\csname l@#1\endcsname
\fi
#2}}
\providecommand{\BIBdecl}{\relax}
\BIBdecl

\bibitem{rockstrom2009planetary}
J.~Rockstrom, W.~Steffen, K.~Noone, A.~Persson, F.~S. Chapin, E.~F. Lambin, T.~M. Lenton, M.~Scheffer, C.~Folke, H.~J. Schellnhuber \emph{et~al.}, ``A safe operating space for humanity,'' \emph{Nature}, vol. 461, no. 7263, pp. 472--475, 2009.

\bibitem{steffen2015planetary}
W.~Steffen, K.~Richardson, J.~Rockström, S.~E. Cornell, I.~Fetzer, E.~M. Bennett, R.~Biggs, S.~R. Carpenter, W.~de~Vries, C.~A. de~Wit \emph{et~al.}, ``Planetary boundaries: Guiding human development on a changing planet,'' \emph{Science}, vol. 347, no. 6223, p. 1259855, 2015.

\bibitem{maier1998architecting}
M.~W. Maier, ``Architecting principles for systems-of-systems,'' \emph{Systems Engineering}, vol.~1, no.~4, pp. 267--284, 1998.

\bibitem{jamshidi2017systems}
M.~Jamshidi, \emph{Systems of Systems Engineering: Principles and Applications}.\hskip 1em plus 0.5em minus 0.4em\relax CRC press, 2017.

\bibitem{do2021context}
L.~V. do~Nascimento, G.~M. Machado, V.~Maran, and J.~P.~M. de~Oliveira, ``Context recognition and ubiquitous computing in smart cities: a systematic mapping,'' \emph{Computing}, vol. 103, no.~5, pp. 801--825, 2021.

\bibitem{kitchenham2007guidelines}
B.~Kitchenham and S.~Charters, ``Guidelines for performing systematic literature reviews in software engineering,'' \emph{EBSE Technical Report}, 2007.

\bibitem{lytras2023digital}
M.~D. Lytras, B.~Alsaywid, and A.~Housawi, ``Digital transformation and smart cities: Insights from the healthcare domain,'' in \emph{Smart Cities and Digital Transformation: Empowering Communities, Limitless Innovation, Sustainable Development and the Next Generation}.\hskip 1em plus 0.5em minus 0.4em\relax Emerald Publishing Limited, 2023, pp. 319--325.

\bibitem{lee2015cyber}
J.~Lee, B.~Bagheri, and H.-A. Kao, ``A cyber-physical systems architecture for industry 4.0-based manufacturing systems,'' \emph{Manufacturing Letters}, vol.~3, pp. 18--23, 2015.

\bibitem{bacco2017environmental}
M.~Bacco, F.~Delmastro, E.~Ferro, and A.~Gotta, ``Environmental monitoring for smart cities,'' \emph{IEEE Sensors Journal}, vol.~17, no.~23, pp. 7767--7774, 2017.

\bibitem{fang2012smart}
X.~Fang and et~al., ``Smart grid—the new and improved power grid: A survey,'' \emph{IEEE Communications Surveys \& Tutorials}, vol.~14, no.~4, pp. 944--980, 2012.

\bibitem{bera2025advancing}
M.~Bera, S.~Das, S.~Garai, S.~Dutta, M.~R. Choudhury, S.~Tripathi, and G.~Chatterjee, ``Advancing energy efficiency: innovative technologies and strategic measures for achieving net zero emissions,'' \emph{Carbon Footprints}, vol.~4, no.~1, pp. N--A, 2025.

\bibitem{adler2024engineering}
R.~Adler, F.~Elberzhager, and F.~Balduf, ``Engineering a sustainable world by enhancing the scope of systems of systems engineering and mastering dynamics,'' \emph{arXiv preprint arXiv:2401.14047}, 2024.

\bibitem{david2024circular}
I.~David, D.~Bork, and G.~Kappel, ``Circular systems engineering,'' \emph{Software and Systems Modeling}, vol.~23, no.~2, pp. 269--283, 2024.

\bibitem{fang2011smart}
X.~Fang, S.~Misra, G.~Xue, and D.~Yang, ``Smart grid—the new and improved power grid: A survey,'' \emph{IEEE communications surveys \& tutorials}, vol.~14, no.~4, pp. 944--980, 2011.

\bibitem{chabot2013small}
D.~Chabot and D.~M. Bird, ``Small unmanned aircraft: precise and convenient new tools for surveying wetlands,'' \emph{Journal of Unmanned Vehicle Systems}, vol.~1, no.~01, pp. 15--24, 2013.

\bibitem{hodgson2016best}
J.~C. Hodgson and L.~P. Koh, ``Best practice for minimising unmanned aerial vehicle disturbance to wildlife in biological field research,'' \emph{Current Biology}, vol.~26, no.~10, pp. R404--R405, 2016.

\bibitem{basili_weiss}
V.~R. Basili and D.~M. Weiss, ``A methodology for collecting valid software engineering data,'' in \emph{IEEE Transactions on Software Engineering}, vol.~10, no.~6, 1984, pp. 728--738.

\bibitem{webster2002analyzing}
J.~Webster and R.~T. Watson, ``Analyzing the past to prepare for the future: Writing a literature review,'' \emph{MIS quarterly}, vol.~26, no.~2, pp. 13--23, 2002.

\bibitem{guessi2015systematic}
M.~Guessi, V.~V. Neto, T.~Bianchi, K.~R. Felizardo, F.~Oquendo, and E.~Y. Nakagawa, ``A systematic literature review on the description of software architectures for systems of systems,'' in \emph{Proceedings of the 30th Annual ACM SAC}, 2015, pp. 1433--1440.

\bibitem{petersen2015guidelines}
K.~Petersen, R.~Feldt, S.~Mujtaba, and M.~Mattsson, ``Guidelines for conducting systematic mapping studies in software engineering: An update,'' \emph{Information and Software Technology}, vol.~64, pp. 1--18, 2015.

\bibitem{hipel2013tackling}
K.~W. Hipel, ``Tackling climate change: A system of systems engineering perspective a research seminar by,'' in \emph{2013 ICSSE}.\hskip 1em plus 0.5em minus 0.4em\relax IEEE, 2013, pp. 11--12.

\bibitem{hall2014water}
P.~Hall, ``Water, energy and food security: Applying wiener's theories to validate a macro-economic nexus model,'' in \emph{2014 IEEE 21CW}.\hskip 1em plus 0.5em minus 0.4em\relax IEEE, 2014, pp. 1--8.

\bibitem{liu2016reengineering}
S.~Liu, K.~P. Triantis, and J.~Xu, ``Reengineering urban operations management and administration by constructing and using urban hierarchical vulnerability indices: An implication of system of systems and big data,'' in \emph{2016 Annual IEEE SysCon}.\hskip 1em plus 0.5em minus 0.4em\relax IEEE, 2016, pp. 1--6.

\bibitem{fritz2008conceptual}
S.~Fritz, R.~J. Scholes, M.~Obersteiner, J.~Bouma, and B.~Reyers, ``A conceptual framework for assessing the benefits of a global earth observation system of systems,'' \emph{IEEE Systems journal}, vol.~2, no.~3, pp. 338--348, 2008.

\bibitem{geldenhuys2018literature}
H.~Geldenhuys, A.~Brent, and I.~De~Kock, ``Literature review for infrastructure transition management towards smart sustainable cities,'' in \emph{2018 IEEE ISSE}.\hskip 1em plus 0.5em minus 0.4em\relax IEEE, 2018, pp. 1--7.

\bibitem{teymourifar2023using}
A.~Teymourifar and M.~A. Trindade, ``Using dematel and ism for designing green public policies based on the system of systems approach,'' \emph{Sustainability}, vol.~15, no.~14, p. 10765, 2023.

\bibitem{sebestyen2021applicability}
V.~Sebesty{\'e}n, T.~Czvetk{\'o}, and J.~Abonyi, ``The applicability of big data in climate change research: The importance of system of systems thinking,'' \emph{Frontiers in Environmental Science}, vol.~9, p. 619092, 2021.

\bibitem{sheridan2024contextualising}
C.~Sheridan, A.~Thatcher, T.-L. Field, D.~Hildebrandt, M.~Kidd, S.~Nadan, L.~Petrik, and J.~Topkin, ``Contextualising urban sanitation solutions through complex systems thinking: A case study of the south african sanitation system,'' in \emph{2024 AIChE Annual Meeting}.\hskip 1em plus 0.5em minus 0.4em\relax AIChE, 2024.

\bibitem{pastor2024r}
R.~Pastor, A.~Lecuona, J.~P. Cort{\'e}s, D.~Caballero, and A.~Fraga, ``(r-issues) rural interoperable system of systems for unified environmental stewardship,'' \emph{Applied Sciences}, vol.~14, no.~18, p. 8245, 2024.

\bibitem{boldrini2022multi}
E.~Boldrini, S.~Nativi, S.~Pecora, I.~Chernov, and P.~Mazzetti, ``Multi-scale hydrological system-of-systems realized through whos: the brokering framework,'' \emph{International Journal of Digital Earth}, vol.~15, no.~1, pp. 1259--1289, 2022.

\bibitem{thatcher2024applying}
A.~Thatcher, G.~S. Metson, and M.~Sepeng, ``Applying the sustainable system-of-systems framework: wastewater (s) in a rapidly urbanising south african settlement,'' \emph{Ergonomics}, vol.~67, no.~4, pp. 450--466, 2024.

\bibitem{yap2024application}
T.~L. Yap, N.~T. Vu, and P.~H. Yeow, ``Application of the sustainable system-of-systems approach and econometric analysis to address china’s decarbonisation problem,'' \emph{Ergonomics}, vol.~67, no.~4, pp. 482--497, 2024.

\bibitem{heitmann2019requirements}
F.~Heitmann, C.~Pahl-Wostl, and S.~Engel, ``Requirements based design of environmental system of systems: Development and application of a nexus design framework,'' \emph{sustainability}, vol.~11, no.~12, p. 3464, 2019.

\bibitem{guan2023scalable}
K.~Guan, Z.~Jin, B.~Peng, J.~Tang, E.~H. DeLucia, P.~C. West, C.~Jiang, S.~Wang, T.~Kim, W.~Zhou \emph{et~al.}, ``A scalable framework for quantifying field-level agricultural carbon outcomes,'' \emph{Earth-Science Reviews}, vol. 243, p. 104462, 2023.

\bibitem{arnautovic2012value}
E.~Arnautovic and D.~Svetinovic, ``Value models for engineering of complex sustainable systems,'' \emph{Procedia Computer Science}, vol.~8, pp. 53--58, 2012.

\bibitem{leach2019liveable}
J.~M. Leach, C.~D. Rogers, A.~Ortegon-Sanchez, and N.~Tyler, ``The liveable cities method: establishing the case for transformative change for a uk metro,'' in \emph{Proceedings of the Institution of Civil Engineers-Engineering Sustainability}, vol. 173, no.~1.\hskip 1em plus 0.5em minus 0.4em\relax Thomas Telford Ltd, 2019, pp. 8--19.

\bibitem{lai2022urban}
Y.~Lai, ``Urban intelligence for carbon neutral cities: creating synergy among data, analytics, and climate actions,'' \emph{Sustainability}, vol.~14, no.~12, p. 7286, 2022.

\bibitem{vercruysse2019interoperability}
K.~Vercruysse, D.~A. Dawson, and N.~Wright, ``Interoperability: A conceptual framework to bridge the gap between multifunctional and multisystem urban flood management,'' \emph{Journal of Flood Risk Management}, vol.~12, no.~S2, p. e12535, 2019.

\bibitem{delgado2019big}
J.~Delgado, N.~Short, D.~Roberts, and B.~Vandenberg, ``Big data analysis for sustainable agriculture on a geospatial cloud framework. frontiers in sustainable food systems 3,'' 2019.

\bibitem{elnashai2021vision}
A.~Elnashai and H.~Mahmoud, ``A vision for smart and sustainable cities,'' pp. 185--188, 2021.

\bibitem{howell2017towards}
S.~Howell, Y.~Rezgui, J.-L. Hippolyte, B.~Jayan, and H.~Li, ``Towards the next generation of smart grids: Semantic and holonic multi-agent management of distributed energy resources,'' \emph{Renewable and Sustainable Energy Reviews}, vol.~77, pp. 193--214, 2017.

\bibitem{rogers2023delivering}
C.~D. Rogers, N.~Grayson, J.~P. Sadler, L.~Chapman, C.~J. Bouch, M.~Cavada, and J.~M. Leach, ``Delivering sustainable, resilient and liveable cities via transformed governance,'' \emph{Frontiers in Sustainable Cities}, vol.~5, p. 1171996, 2023.

\bibitem{chang2024climate}
C.~M. Chang and A.~Hossain, ``A climate adaptation asset risk management approach for resilient roadway infrastructure,'' \emph{Infrastructures}, vol.~9, no.~12, p. 226, 2024.

\bibitem{beckett2023pathway}
R.~C. Beckett and J.~Mo, ``The pathway to sustainability in transdisciplinary system development,'' in \emph{Leveraging Transdisciplinary Engineering in a Changing and Connected World}.\hskip 1em plus 0.5em minus 0.4em\relax IOS Press, 2023, pp. 863--872.

\bibitem{albaaji2023artificial}
G.~F. Albaaji and V.~C. SS, ``Artificial intelligence sos framework for sustainable agricultural production,'' \emph{Computers and Electronics in Agriculture}, vol. 213, p. 108182, 2023.

\bibitem{prakashacolossus}
P.~S. Prakasha, N.~Naeem, K.~Amadori, G.~Donelli, J.~Akbari, F.~Nicolosi, L.~K. Franz{\'e}n, M.~Ruocco, T.~Lefebvre, and B.~Nagel, ``Colossus eu project--collaborative sos exploration of aviation products, services and business models: Overview and approach,'' in \emph{Proc. of 34th Congress of the International Council of the Aeronautical Sciences}, 2024.

\bibitem{markey2023managing}
C.~L. Markey and S.~Ahmed-Kristensen, ``Managing smart systems for the net zero agenda--how can digital twin technologies and smart products deliver customer value?'' \emph{Proceedings of the Design Society}, vol.~3, pp. 2545--2554, 2023.

\bibitem{nozarian2024system}
M.~Nozarian, A.~Fereidunian, and M.~Barati, ``A system of systems approach to reliability-oriented planning of people-centric smart city energy infrastructure: A bilevel milp formulation,'' \emph{IEEE Systems Journal}, 2024.

\bibitem{abdoli2021analyzing}
S.~Abdoli and B.~Kianian, ``Analyzing the environmental consequences of production processes from a system of systems perspective: a case of gear manufacturing in the automotive industry,'' \emph{Procedia CIRP}, vol.~98, pp. 376--381, 2021.

\bibitem{watson2020systems}
B.~C. Watson, M.~J. Weissburg, and B.~Bras, ``Systems of systems engineering to improve resilience: A case study comparison of biologically inspired and traditional approaches,'' in \emph{International Design Engineering Technical Conferences and Computers and Information in Engineering Conference}, vol. 83976.\hskip 1em plus 0.5em minus 0.4em\relax American Society of Mechanical Engineers, 2020, p. V008T08A020.

\bibitem{little2019tiered}
J.~C. Little, E.~T. Hester, S.~Elsawah, G.~M. Filz, A.~Sandu, C.~C. Carey, T.~Iwanaga, and A.~J. Jakeman, ``A tiered, system-of-systems modeling framework for resolving complex socio-environmental policy issues,'' \emph{Environmental Modelling \& Software}, vol. 112, pp. 82--94, 2019.

\bibitem{tariq2013determinants}
Z.~Tariq, S.~Cavalieri, and R.~Pinto, ``Determinants of smart energy demand management: An exploratory analysis,'' in \emph{Advances in Production Management Systems. Sustainable Production and Service Supply Chains: IFIP WG 5.7 International Conference, APMS 2013, State College, PA, USA, September 9-12, 2013, Proceedings, Part II}.\hskip 1em plus 0.5em minus 0.4em\relax Springer, 2013, pp. 548--555.

\bibitem{hoffmann2014improved}
A.~Hoffmann, J.~Penner, K.~Vohland, W.~Cramer, R.~Doubleday, K.~Henle, U.~K{\~o}ljalg, I.~K{\"u}hn, W.~E. Kunin, J.~J. Negro \emph{et~al.}, ``Improved access to integrated biodiversity data for science, practice, and policy-the european biodiversity observation network (eu bon),'' \emph{Nature Conservation}, no.~6, pp. 49--66, 2014.

\bibitem{hernandez2014multi}
E.~A. Hernandez, V.~Uddameri, and M.~A. Arreola, ``A multi-media planning model for assessing co-located energy and desalination plants,'' \emph{Environmental earth sciences}, vol.~71, pp. 2673--2686, 2014.

\bibitem{phillis2012system2}
Y.~A. Phillis and V.~S. Kouikoglou, ``System-of-systems hierarchy of biodiversity conservation problems,'' \emph{Ecological Modelling}, vol. 235, pp. 36--48, 2012.

\bibitem{sandor2008transitioning}
D.~Sandor and C.~Riley, ``Transitioning to biofuels: A system-of-systems perspective,'' National Renewable Energy Lab.(NREL), Golden, CO (United States), Tech. Rep., 2008.

\bibitem{scholes2012building}
R.~J. Scholes, M.~Walters, E.~Turak, H.~Saarenmaa, C.~H. Heip, {\'E}.~{\'O}. Tuama, D.~P. Faith, H.~A. Mooney, S.~Ferrier, R.~H. Jongman \emph{et~al.}, ``Building a global observing system for biodiversity,'' \emph{Current Opinion in Environmental Sustainability}, vol.~4, no.~1, pp. 139--146, 2012.

\bibitem{phillis2012system}
Y.~A. Phillis and V.~S. Kouikoglou, ``A system-of-systems approach to the analysis and conservation of biodiversity,'' in \emph{Enterprise Information Systems: 13th International Conference, ICEIS 2011, Beijing, China, June 8-11, 2011, Revised Selected Papers 13}.\hskip 1em plus 0.5em minus 0.4em\relax Springer, 2012, pp. 3--15.

\bibitem{pruyt2007transition}
E.~Pruyt and W.~Thissen, ``Transition of the european electricity system and system of systems concepts,'' in \emph{2007 IEEE SOSE}.\hskip 1em plus 0.5em minus 0.4em\relax IEEE, 2007, pp. 1--6.

\bibitem{riley2008transitioning}
C.~Riley and D.~Sandor, ``Transitioning to biofuels: A system-of-systems perspective,'' National Renewable Energy Lab.(NREL), Golden, CO (United States), Tech. Rep., 2008.

\bibitem{pruyt2011energy}
E.~Pruyt, J.~Kwakkel, G.~Yucel, and C.~Hamarat, ``Energy transitions towards sustainability: A staged exploration of complexity and deep uncertainty,'' in \emph{Proceedings of the 29th International Conference of the System Dynamics Society}, 2011, pp. 1--26.

\bibitem{gomes2022flood}
M.~N. Gomes~J{\'u}nior, M.~H. Giacomoni, A.~F. Taha, and E.~M. Mendiondo, ``Flood risk mitigation and valve control in stormwater systems: State-space modeling, control algorithms, and case studies,'' \emph{Journal of Water Resources Planning and Management}, vol. 148, no.~12, p. 04022067, 2022.

\bibitem{manzano2020dynamic}
W.~Manzano, V.~V. Graciano~Neto, and E.~Y. Nakagawa, ``Dynamic-sos: An approach for the simulation of systems-of-systems dynamic architectures,'' \emph{The Computer Journal}, vol.~63, no.~5, pp. 709--731, 2020.

\end{thebibliography}

\end{document}